# Einstein's cosmic model of 1931: an analysis and translation of a forgotten model of the universe

C. O'Raifeartaigh and B. McCann

*Department of Computing, Maths and Physics, Waterford Institute of Technology, Cork Road, Waterford, Ireland*

Author for correspondence: coraifeartaigh@wit.ie

## Abstract

We present an analysis and translation of Einstein's 1931 paper "*Zum kosmologischen Problem der allgemeinen Relativitätstheorie*" or "*On the cosmological problem of the general theory of relativity*". In this little-known paper, Einstein proposes a cosmic model in which the universe undergoes an expansion followed by a contraction, quite different to the monotonically expanding Einstein-de Sitter model of 1932. The paper offers many insights into Einstein's cosmology in the light of the first evidence for an expanding universe and we consider his views of issues such as the curvature of space, the cosmological constant, the singularity and the timespan of the expansion.

A number of original findings concerning Einstein's 1931 model are presented. We find that Einstein's calculations of the present radius and matter density of the universe contain some anomalies: we suggest that his estimate of the age of the universe is based on a questionable calculation of Friedmann's: we find that the model is not periodic, contrary to what is often stated. A first English translation of Einstein's paper is included as an appendix.

## 1. Introduction

Einstein's paper "*Zum kosmologischen Problem der allgemeinen Relativitätstheorie*" (Einstein 1931) is of historical interest because it proposes a model of the universe distinct from his static model of 1917 (Einstein 1917) or the monotonic Einstein-de Sitter model of 1932 (Einstein and de Sitter 1932). The paper marks the first scientific publication in which Einstein formally rejects the notion of a static universe and explores the possibility of a cosmos of time-varying radius.

With the publication of Hubble's discovery of an approximately linear relation between the recession of the spiral nebulae and their distance (Hubble 1929), several physicists began to consider the possibility of a universe of expanding radius. Thus, Einstein's paper can be viewed in the context of a number of works on relativistic cosmology such as those by Lemaître (1927, 1931), Eddington (1930a, b), de Sitter (1930a, b) and Tolman (1929, 1930). The distinguishing feature of Einstein's contribution is that he is keen to investigate whether a relativistic model can account for the new observations if the cosmological constant is removed from the field equations. Adopting Friedmann's analysis of a relativistic universe of spherical curvature and time-varying radius (Friedmann 1922), Einstein sets the cosmological constant to zero and arrives at a model of a universe that first expands and then contracts, with singularity-like behaviour at either end.[1] As one of a very small number of papers he published on cosmology, the article offers many insights into Einstein's view of emerging puzzles such as the singularity and the timespan of the expansion.

It is known that Einstein's paper was written over the course of four days in April 1931, following his return to Berlin after a three-month stay in the United States.[2] Much of this trip was spent at Caltech in Pasadena, where Einstein had many interactions with the relativist Richard Tolman. The trip also included a meeting with Edwin Hubble and other astronomers at the Mt Wilson Observatory.[3] Press reports of seminars given by Einstein in Pasadena suggest that he was willing to view Hubble's observations of a redshift/distance relation for the spiral nebulae as possible evidence for an expanding universe. For example, the *New York Times* reported Einstein as commenting that *"New observations by Hubble and Humason concerning the redshift of light in distant nebulae make the presumptions near that the general structure of the universe is not static*" (AP 1931a) and "*The redshift of the distant nebulae have smashed my old construction like a hammer blow"* (AP 1931b).[4] A more detailed discussion of Einstein's reaction to Hubble's observations can be found in the essay by Harry Nussbaumer in this volume (Nussbaumer 2014).

---

[1] The model is sometimes known as the Friedmann-Einstein universe, see (North 196, p132) and (Rindler 1969 , p261).

[2] This is annotated in Einstein's diary and discussed in several accounts such as (Eisinger 2012, chapter 7).

[3] Accounts of Einstein's time in Pasadena can be found in (Bartusiak 2009, chapter 16), (Nussbaumer and Bieri 2013, chapter 14) and (Eisinger 2012, chapter 7).

We present a guided tour of Einstein's 1931 paper in section 2 of this report. Einstein's view of particular issues such as the instability of his static solution, the cosmological constant, the curvature of space, the singularity and the cyclic universe are then considered in detail in sections 3 to 7 respectively. In section 8, we review Einstein's use of Hubble's observations to extract estimates for the present radius and matter density of the universe from his model and find that his calculations contain a numerical error; this finding is supported by writing on a blackboard used by Einstein during a lecture at Oxford University in May 1931. In section 9, we consider Einstein's view of the problem of the age of the universe, while a more general discussion of his philosophy of cosmology is provided in section 10. A full English translation of Einstein's 1931 paper is provided in the appendix.[5]

## 2. A guided tour of the paper

Einstein begins his paper by defining what he means by the cosmological problem:

> "*The cosmological problem is understood to concern the question of the nature of space and the manner of the distribution of matter on a large scale, where the material of the stars and stellar systems is assumed for simplicity to be replaced by a continuous distribution of matter*".

He remarks that numerous theoretical articles on cosmology have appeared in the literature since his original 1917 paper on relativistic cosmology, and that Hubble's observations suggest new directions for theorists to consider:

[4] These are journalists' reports of what Einstein said and may not be his exact words.
[5] As Einstein's paper is not available in English translation, we provide our own translation for the purposes of this article by kind permission of the Albert Einstein Archives of the Hebrew University of Jerusalem.

"*Since I began to address this problem, shortly after advancing the general theory of relativity, not only have numerous theoretical articles on the subject appeared, but facts have come to light from Hubbel's* [sic] *investigations of the Doppler effect and the distribution of extra-galactic nebulae that indicate new directions for the theory".*

We note that this sentence does not contain any specific references to seminal works on relativistic cosmology such as the papers of de Sitter, Weyl or Robertson (de Sitter 1917; Weyl 1922; Roberston 1929). We also note that there is also no specific reference to Hubble's paper of 1929 (Hubble 1929) and that the latter's name is misspelt throughout the paper, suggesting that Einstein is not very familiar with Hubble's work.[6]

Einstein then provides a brief review of his static cosmic model of 1917, recalling his key assumptions, the introduction of the cosmological constant, and his solution for the present world radius:

*"In my original investigation, I proceeded from the following assumptions:*

1. *All locations in the universe are equivalent; in particular the locally averaged density of stellar matter should therefore be the same everywhere.*
2. *Spatial structure and density should be constant over time.*

*At that time, I showed that both assumptions can be accounted for with a non-zero mean density ρ, if the so-called cosmological term is introduced into the field equations of the general theory of relativity such that these read:*

$$\left(R_{ik} - \frac{1}{2} g_{ik} R\right) + \lambda g_{ik} = -\kappa T_{ik} \qquad \ldots\ldots \quad (1)$$

*These equations can be satisfied by a spatially spherical static world of radius* $P = \sqrt{\frac{2}{\kappa\rho}}$ *where ρ is the (pressure-free) mean density of matter".*

He then points out that Hubble's observations may have rendered the assumption of a static universe invalid and asks whether relativity can account for the new findings:

[6] This point is discussed in more detail in the essay by Harry Nussbaumer in this issue

*"Now that it has become clear from Hubbel's* [sic] *results that the extra-galactic nebulae are uniformly distributed throughout space and are in dilatory motion (at least if their systematic redshifts are to be interpreted as Doppler effects), assumption (2) concerning the static nature of space has no longer any justification and the question arises as to whether the general theory of relativity can account for these findings."*.

He notes that some theoretical attempts have already been made to explain the new observations:

*"Several investigators have attempted to account for the new facts by means of a spherical space, whose radius P is variable over time*".

Once again, no specific citations are made, so we can only presume that Einstein is referring to works such as those by Lemaître, Eddington, de Sitter and Tolman (Lemaître 1927; Eddington 1930; de Sitter 1930a, b; Tolman 1929, 1930). Indeed, the only specific reference in the entire paper is to Alexander Friedmann's model of 1922:

"*The first to try this approach, uninfluenced by observations, was A. Friedman, on whose calculations I base the following remarks".*

With the introduction over, Einstein sets about constructing his new cosmic model with the help of Friedmann's analysis:

*"One proceeds accordingly from a line element of the form*

$$ds^2 = -P^2(dx_1^2 + sin^2 x_1\, dx_2^2 + sin^2 x_1 sin^2 x_2 dx_3^2) + c^2 dx_4^2 \quad .... (2)$$

*where P is understood to be a function of the real-valued time variable $x_4$ alone. For the determination of P and the relation of this quantity to the (variable) density ρ he derives from (1) the two differential equations*

$$\frac{P'^2}{P^2} + \frac{2P''}{P} + \frac{c^2}{P^2} - \lambda = 0 \quad .... \qquad (2)$$

$$\frac{3P'^2}{P^2} + \frac{3c^2}{P^2} - \lambda = \kappa c^2 \rho \quad .... \qquad (3)$$

Einstein notes first that equations (2) and (3) above reduce to his former solution if a static world radius is assumed:

> *"From these equations, one obtains my previous solution by assuming that P is constant over time."*

He then comments that the static solution is unstable:

> "*However, it can also be shown with the help of these equations that this solution is not stable, i.e., that a solution that deviates only slightly from the static solution at a given point in time will differ ever more from it with the passage of time".*

It is not clear whether Einstein reaches this conclusion by inspection of equations (2) and (3) above, or by reference to earlier work by Eddington, as discussed in section 3. In any event, Einstein concludes that his former solution is invalid, irrespective of astronomical observations:

> *"On these grounds alone, I am no longer inclined to ascribe a physical meaning to my former solution, quite apart from Hubbel's* [sic] *observations*".

Einstein then proceeds to explore whether relativity can account for the new observations without the cosmological constant:

> "*Under these circumstances, one must ask whether one can account for the facts without the introduction of the λ-term, which is in any case theoretically unsatisfactory".*

Setting λ = 0, he integrates equation (2) to obtain:

$$(\frac{dP}{dt})^2 = c^2 \frac{P_0 - P}{P} \text{ , ... ....} \qquad (2a)$$

He notes that the integration constant $P_0$ represents a maximum bound for the cosmic radius, which is also a turning point:

> "*...where $P_0$ denotes a constant of integration that sets an upper bound for the world radius that cannot be exceeded with the passage of time. At this point a change of sign for $\frac{dP}{dt}$ must occur.*"

Einstein then invokes Hubble's observations, commenting that they imply that the cosmic radius is currently increasing, and that they give a measure of the quantity $\frac{1}{P}\frac{dP}{dt}\frac{1}{c}$ which he labels *D*:

> "*It follows from Hubbel's results that for the present, it is to be assumed that $\frac{dP}{dt} > 0$, and that the Doppler effect divided by the distance is a quantity that is independent of distance, which here can be expressed with sufficient accuracy by the quantity $D = \frac{1}{P}\frac{dP}{dt}.\frac{1}{c}$.*"

Thus, he rewrites equations (2a) and (3) in terms of the observational parameter *D,* obtaining the equations

$$D^2 = \frac{1}{P^2}\frac{P_0 - P}{P} \text{ ...} \qquad (2b)$$

$$D^2 = \frac{1}{3}\kappa\rho\frac{P_0 - P}{P_0} \text{ ....} \qquad (3a)$$

Einstein then gives a more detailed description of the evolution of the radius over time:

*"For small P (our idealization fails for the strict limiting case P = 0), P grows very rapidly. Thereafter the rate of change $\frac{dP}{dt}$ decreases ever more with increasing P, and disappears when the limit $P = P_0$ is attained, whereupon the whole process is gone through in the opposite sense (i.e. with ever rapidly decreasing P)."*

We note the comment in parenthesis concerning the failure of the model at the point $P = 0$; this comment is discussed in section 6 below.

In the next section of the paper, Einstein extracts estimates for the density of matter and the present world radius from his model. Assuming that the current world radius is of the same order of magnitude as the maximum value, he simplifies equations (3a) and (2b) to obtain

$$D^2 \sim \kappa\rho \qquad \text{and} \qquad P \sim 1/D$$

Substituting for *D* from Hubble's observations, Einstein obtains estimates of $10^{-26}$ g/cm$^3$ and $10^8$ light-years for the present matter density $\rho$ and world radius *P* respectively. We suggest that a numerical error has occurred in these calculations, as described in section 8 below.

In the final part of the paper, Einstein remarks that the timespan of the expansion presents a significant difficulty for the model:

"*... the greatest difficulty with the whole approach, as is well-known, is that according to (2 a), the elapsed time since P = 0 comes out at only about $10^{10}$ years."*

The nature of the "difficulty" is not spelt out, but the remark probably refers to estimates of stellar ages from astrophysics, as discussed in section 9 below. It is also not clear how Einstein obtains an age estimate of ten billion years from equation (2a); we suggest that the figure is taken from a rough calculation of Friedmann's, as discussed in section 9.

We note that Einstein considers the elapsed time a problem for all expanding models:

*"Moreover, it should be noted that it is unlikely that any theory that interprets Hubbel's tremendous shifts of spectral lines as Doppler effects will easily avoid this difficulty*."

He suggests that the problem may be a consequence of the assumption of homogeneity in the models:

> *"One can seek to escape this difficulty by noting that the inhomogeneity of the distribution of stellar material makes our approximate treatment illusory".*

This view is discussed in section 9.

Einstein concludes the paper by remarking that his new cosmic model is simple enough to compare with observations, and that a model without the cosmological constant can account for the new observations more naturally than the static model.

> "*In any case, this theory is simple enough that it can be easily compared with the astronomical facts…most importantly, it seems that the general theory of relativity can account for Hubbel's new facts more naturally (namely, without the λ-term), than it can the postulate of the quasi-static nature of space, which has now been rendered a remote possibility empirically."*

## 3. On the instability of the static universe

Einstein's starting point for his new model is the instability of his former static solution. This is of great importance as it renders a universe of time-independent radius unlikely, irrespective of astronomical observations. (It was not clear at this time whether Hubble's observations constituted conclusive evidence for an expanding universe, as discussed below). Thus it is surprising that Einstein merely states that "*it can also be shown with the help of these equations that this solution is not stable"* but does not demonstrate precisely how he comes to this conclusion. The statement could be the observation that the derivatives in equations (2) and (3) become non-zero if $P$ deviates slightly from a constant value; however, Einstein does not suggest any physical mechanism for the initial variation of $P$.

By contrast, the question of instability was considered in some detail by Eddington the year before (Eddington 1930). Considering differential equations similar to equations (2) and (3) above, Eddington noted that a slight decrease in matter density would cause a small expansion, resulting in a further decrease in density (with a similar feedback effect for an increase in density). Indeed, this observation of the instability of static solutions convinced Eddington to take dynamic models seriously: "*The proof of the instability of Einstein's model greatly strengthens our grounds for interpreting the recession of the spiral nebulae as an indication of world curvature*" (Eddington 1930).

Thus, it is possible that Einstein's statement above on the instability of his static solution is a reference to Eddington's paper of 1930. Certainly, it is likely that Einstein was aware of the work, given his visit to Eddington in the summer of that year (Vibert Douglas 1952, p102).

It is interesting to ask why the instability of his static solution was not apparent to Einstein in 1917. One answer may be that the instability is more easily seen from the differential equations (2) and (3) above, equations that only arise if one allows for the possibility of time-varying solutions for the world radius. Thus Einstein's original static universe is a good example of the risks of *a priori* assumptions,[7] while Friedmann, *"uninfluenced by observations",* allowed for all possibilities.[8]

## 4. On the cosmological constant

As is well-known, Einstein introduced the cosmological constant to the field equations of general relativity because of his assumption of a universe of "*spatial structure and density that was constant over time*" (see section 2). In the absence of any observational evidence known to him for a dynamic universe,[9] he added the term in order to counterbalance the attraction of gravity, giving a universe of cylindrical curvature that was static in time. Thus the 'cosmological constant' allowed Einstein to postulate a finite universe of closed spatial geometry whose radius could be calculated from the density of matter.[10]

---

[7] Although Einstein's assumption of a static universe in 1917 was reasonable given the absence of any evidence to the contrary known to him, the assumption was more philosophical than empirical as there was no a priori guarantee that an expansion of space on the largest scales would be observable by astronomy.

[8] A recent review of Friedmann's analysis can be found in (Belenkiy 2012, 2013).

[9] It could be argued that the redshift observations of Vesto Slipher pointed towards a cosmic expansion (Peacock 2013) but there is no evidence Einstein was aware of these observations.

[10] Constants of integration occur naturally in the solution of equations such as the Einstein field equations. The size of the cosmological constant is constrained by the requirement that relativity predict the motion of the planets, but there is no reason it should be exactly zero. Einstein's original suggestion was that a small, non-zero constant of integration could simultaneously render the universe static and give it a closed curvature, neatly removing the problem of boundary conditions.

There is no doubt that Einstein himself saw the term as something of an ad-hoc addition, remarking at the end of the 1917 paper that "*In order to arrive at this consistent view, we admittedly had to introduce an extension of the field equations of gravity which is not justified by our actual knowledge of gravitation. …that term is necessary only for the purpose of making possible a quasi-static distribution of matter, as required by the fact of the small velocities of the stars"* (Einstein 1917). As early as 1923, following Hermann Weyl's demonstration (Weyl 1923) of an apparent recession of test particles in the de Sitter model of 1917 (de Sitter 1917), Einstein remarked "*If there is no quasi-static world, then away with the cosmological term*" (Einstein 1923).

It is therefore no surprise that Einstein is willing to jettison the cosmological constant in the face of emerging evidence for a non-static universe. As mentioned in our introduction, it is clear from his reported comments in early 1931 that he viewed Hubble's observations as possible evidence for a universe that was expanding on the largest scales (albeit with some reservations, as discussed in section 10 below). Hence, Einstein's strategy is to investigate how well a simple relativistic model, namely one without a cosmological constant, can account for the facts.

Thus Einstein removes the entire term $\lambda g_{ik}$ from the field equations, formally justifying its removal on both theoretical and observational grounds. As discussed in section 3 above, the theoretical justification is that his original static solution was in any case unstable. The experimental justification is that Einstein suspects that a model without a cosmological constant will give a good match to the observational data. Thus he constructs a new cosmic model by setting with λ = 0 in Friedmann's 1922 analysis, tests it by extracting estimates for the current mean density of matter and the radius of the universe, and concludes *"..it seems that the general theory of relativity can account for Hubbel's* [sic] *new facts more naturally (namely, without the λ-term), than it can the postulate of the quasi-static nature of space, which has now been empirically rendered a remote possibility."*

As evidence for an expanding universe strengthened over the years, Einstein was never to reinstate the cosmological constant. His view of the term as an unnecessary addition to the field equations did not waver over the years, as can be seen in statements such as *"Since the expansion of the stellar systems has become known empirically, there are for the present no logical or physical grounds for the introduction of that term"* (Einstein 1934) and *"The introduction of a second member constitutes a complication of the theory, which seriously reduces its logical simplicity"* (Einstein 1945). Indeed, he is said to have considered the term his "greatest blunder".[11] Certainly, the removal of the term in this paper is very much in line with Einstein's minimalist approach to cosmology (see section 10); if the cosmological constant was introduced to keep the universe static, why keep it when the latter assumption is no longer justified by the evidence?

In retrospect, a subtle mathematical flaw can be seen in this reasoning. Einstein's 'blunder' in 1917 was not the introduction of a cosmological constant in principle, since one cannot discount the possibility of an extra term in the field equations (although it must be very small in order for relativity to successfully predict motion on local scales such as the motion of the planets). Instead, the 'blunder' was to assign a specific value to the term for which there was no real justification, thus preventing Einstein from predicting a dynamic universe. It could be argued that Einstein makes the same mistake in the present paper, by once again setting an unknown constant of integration to a particular value, namely zero. Nowadays, there is strong evidence that the expansion of the universe is currently accelerating, a finding that many cosmologists interpret in terms of a non-zero cosmological constant, although the physical nature of the phenomenon remains the subject of much debate.[12]

## 5. On the curvature of space

The assumption of positive spatial curvature in the paper under discussion may seem curious to modern eyes, but it arises primarily from the fact that Einstein works directly from Friedmann's 1922 analysis, which assumed a universe of time-varying radius and spherical curvature (Friedmann 1922). More generally, since relativity replaced the attractive gravitational force of Newton with a curvature of spacetime, positive spatial curvature for a matter-filled universe was commonly assumed at the time (Einstein 1917, de Sitter 1917, Eddington 1923, Eddington 1930, de Sitter 1930). Certainly, it seems that Einstein has not yet considered that a matter-filled universe of expanding radius need not be of such curvature. This impression is confirmed in the opening comment of the later Einstein-deSitter paper *"In a recent note in the Göttinger Nachrichten, Dr. O. Heckmann has pointed out that non-static solutions*

---

[11] Einstein reportedly made his "greatest blunder" comment to the Russian physicist George Gamow (Gamow 1956) and it became extremely well-known. Gamow's account has recently been queried on the basis that there is no record of Einstein making the comment elsewhere (Livio 2013). However, we see no reason to doubt Gamow's account as Einstein undoubtedly disliked the term and Gamow was one of the few German-speaking cosmologists with whom he had contact in later years.

[12] See (Peebles and Ratra 2002) for a review.

*of the field equations of general relativity with constant density do not necessarily imply a positive curvature of three-dimensional space, but that this curvature may be negative or zero"* (Einstein and de Sitter 1932). We note further that it is Heckmann, not Friedmann, who is cited, suggesting that Einstein may not be familiar with Friedmann's exploration of alternative curvatures (Friedmann 1924).

As noted in section 2, setting the cosmological constant to zero in Friedmann's analysis and solving for *dP/dt* leads Einstein to the differential equation

$$\left(\frac{dP}{dt}\right)^2 = c^2\frac{P_0 - P}{P}\ , \dots \dots \qquad (2a)$$

Interpreting Hubble's observations as evidence that the radius is currently increasing, Einstein gives a detailed description from equation (2a) of the time evolution of the radius. "*For small P (our idealization fails for the strictly limiting case P = 0), P grows very rapidly. Thereafter the rate of change $\frac{dP}{dt}$ decreases ever more with increasing P, and disappears when the limit $P = P_0$ is attained, whereupon the whole process is gone through in the opposite sense (i.e. with ever rapidly decreasing P*)." Thus Einstein arrives at a non-monotonic model with a single maximum, i.e., a universe whose radius first increases and then decreases. A graph of this model, reproduced from a later essay by Einstein, is shown in figure 1 (Einstein 1945).

The assumption of positive curvature is the key difference between the paper under discussion and the Einstein-de Sitter model of 1932. In the latter case, the assumption of spatial flatness gives a monotonic solution for *P* (Einstein and de Sitter 1932). Nonetheless, the extraction of estimates for the present matter density and world radius proceeds almost identically. That Einstein and de Sitter calculate values for these quantities that are very different to those calculated by Einstein in the paper under discussion adds to our suspicion that a numerical error occurred in the calculations of this paper, as detailed in section 8.

## 6. On cosmic singularities

The paper under discussion contains a fascinating insight into Einstein's view of the problem of cosmic singularities. As noted in section 2, his approach to the issue is contained in a brief phrase in parentheses: *"..(our idealization fails for the strictly limiting case P = 0)"*. This phrase indicates that Einstein's view is that the model breaks down in the limit *P = 0*.

He does not discount the existence of singularities or speculate about origins, but simply states that the model is not defined for the value $P = 0$ at each end of the cycle.

This attitude is consistent with Einstein's view of non-cosmological singularities in general relativity. He took little interest in the phenomena throughout the 1920s, seeing them as a shortcoming of relativity that might later be overcome in a more complete theory.[13] In any event, it is clear that Einstein views his model as a single evolution over an open interval, i.e., with the initial and endpoints excluded from description. The use of parentheses further indicates that Einstein does not see the problem as one of great physical significance.

This view of singularities is very different to that of Friedmann (Friedmann 1922). The latter viewed the initial and final points as part of the evolutionary cycle, even speculating that each evolution might be followed by another, giving a periodic universe of infinite existence as discussed in the next section.

**7. On periodic worlds and cyclic universes**

The model of this paper has been variously described as 'quasi-periodic' (Tolman 1931, 1932), 'cycloidal' (North 1965, p 132, Rindler 1969, p261), 'oscillatory' (North 1965, p131, Kragh 1996, p 34, Nussbaumer and Bieri 2009, p137), 'periodic' (Steinhardt and Turok 2007, p 176) and 'pulsating' (Kragh 2013). This nomenclature probably arose because the model belongs to a class of solutions originally labelled 'the periodic world' by Friedmann (Friedmann 1922). However, it is important to note that Einstein himself does not use any of these terms in the paper or in later discussions of the model (Einstein 1945). Indeed, we suggest that such terms are somewhat misleading, as discussed below.

For the differential equation (2) above, Friedmann obtained the general solution

$$\frac{1}{c^2}\left(\frac{dP}{dt}\right)^2 = \frac{A - P + \frac{\lambda}{3c^2}P^3}{P} \qquad (5)$$

where $A$ is a constant and we have used Einstein's notation $P$ for the world radius (Friedmann 1922). According to Friedmann, solving for $t$ gives a 'periodic' solution of period

[13] Throughout the 1920s, Einstein viewed the problem of singularities in general relativity as a facet of the incompleteness of the theory (Earman and Eisenstaedt 1999).

$$t_\pi = \frac{2}{c}\int_0^{x_0} \sqrt{\frac{x}{A - x + \frac{\lambda}{3c^2}x^3}}\, dx \quad (6)$$

for values of λ in the range (-∞, $4c^2/9A^2$). Discussing these solutions, he states *"The curvature radius varies thereby between 0 and $x_0$. We want to call this the periodic world".* Thus it seems that Friedmann uses the term 'periodic world' in the sense of a universe whose radius first increases and then decreases. This use of the term does not conform with its standard definition [14] and the matter is further confused by the fact that Friedmann goes on to discuss the possibility of a solution that is genuinely periodic, stating "*..if the time varies between – ∞ and +∞ ...then we arrive at a true periodicity of the space curvature"* (Friedmann 1922). It is important to note that this latter postulate requires some further assumptions since the differential equations (2), (3), (2a) and (5) are not defined for the value $P = 0$.

Einstein's approach is more cautious. While the model of this paper is one of expansion followed by contraction, his attitude is that a breakdown of theory occurs at the endpoints, as discussed in section 6. In consequence, the question of repeat cycles simply does not arise for Einstein.

The theorist Richard Tolman considered time-varying cosmic models of closed curvature such as the one of this paper in some depth (Tolman 1931, 1932). He demonstrated mathematically that such models cannot exhibit a behaviour that is truly periodic, concluding that *"..the sets of requirements for a strictly periodic solution could not be satisfied by any fluid for filling the model of the universe which has reasonable physical properties"* (Tolman 1931). Ascribing this breakdown in theory to the idealized assumptions of the model, Tolman then showed that, if the problem of the singularity was overlooked, "quasi-periodic" models comprising a sequence of expansions and contractions did not violate basic principles of thermodynamics - noting all the while that the analysis *"fails to carry us through the exceptional point of zero volume"* (Tolman 1931).

Taken in conjunction with Friedmann's nomenclature of 'periodic worlds', it is possible that Tolman's authoritive discussion of quasi-periodic models led to retrospective descriptions of the model of this paper as 'cyclic', 'periodic' or 'oscillatory'. However, it is worth emphasizing that Einstein himself did not view his model in this way, for the reasons stated above. Indeed, it is interesting that when the idea of a cyclic universe was discussed by Tolman in Einstein's presence in early 1931, Einstein was quite dismissive of the idea: *"Dr. Tolman suggested that if we had a periodic solution of contraction and expansion it might be satisfactory. Dr. Einstein replied that the equations do not satisfy such a thing, but indicate if such were the case the whole universe might explode, "going swish" he said, laughing.*" (AP 1931b). This comment precedes the paper under discussion but there is nothing in the current paper, or in Einstein's later discussions of the model, to suggest that he changed his opinion on the matter (Einstein 1945). All in all, we suggest that it is not historically or mathematically accurate to describe the model of this paper as periodic, oscillatory or cyclic .

[14] The period of a periodic function is defined as the time to complete one full cycle (among many).

## 8. On Einstein's estimates of the present density of matter and radius of the universe

An important aspect of this paper is that Einstein tests his model by extracting estimates for the current matter density and world radius with the use of Hubble's observations of the recession of the spiral nebulae. These estimates are simple order-of-magnitude calculations intended to reveal whether or not the predictions of the model are in reasonable agreement with estimates of the same quantities from astronomy. However, it is intriguing that the calculations appear to contain a numerical error, as outlined below.

We consider the estimate of the current world radius first, as there is no ambiguity of units in this case. Approximating equation (2b) as $P \sim \frac{1}{D}$ , Einstein obtains a value of $10^8$ light-years ($9.5 \times 10^{25}$ cm) for the current world radius $P$. Although not stated explicitly, this estimate clearly implies that he used a value of $1 \times 10^{-26}$ cm$^{-1}$ for the quantity $D$. Yet we recall from section 2 that the text of the article makes it clear that $D$ is to be obtained from Hubble's observations *"It follows from Hubbel's* [sic] *results that ....the Doppler effect divided by distance is a quantity that is independent of distance which can be expressed with sufficient accuracy from the quantity* $D = \frac{1}{P}\frac{dP}{dt}.\frac{1}{c}$ *"*. Taking the contemporaneous value for the Hubble constant as $H_0$ ~ 500 km s$^{-1}$ Mpc$^{-1}$, [15] we obtain instead

$$D = \frac{500x10^5 cm/s}{(30.85x10^{23} cm)(3x10^{10} cm/s)} = 5.4 \times 10^{-28} \text{ cm}^{-1} ,$$

[15] The term 'Hubble constant' was not used at the time but we use it here for convenience.

a value that is an order of magnitude smaller than that of Einstein.[16] Substituting our value of $D$ into the relation $P \sim \frac{1}{D}$, we obtain an estimate for the present world radius of $P$ = 1.85x10$^{27}$ cm or 2x10$^{9}$ light-years. This estimate of the radius is twenty times larger than that obtained by Einstein in the paper under discussion, but is in close agreement with that (2x10$^{27}$ cm) obtained by Einstein and de Sitter in a similar calculation a year later (Einstein and de Sitter 1932).

As regards his estimate of the density of matter, Einstein approximates equation (3a) as $D^2 \sim \kappa\rho$, and obtains a value for the matter density of $\rho$ = 10$^{-26}$. The units of measurement are not stated but are very likely g/cm$^3$ as these were the units used by Einstein for this quantity in every known instance (Einstein 1917, 1932, 1945). This value is consistent with our suggestion above that Einstein used a value of $D \sim$ 1 x 10$^{-26}$ cm$^{-1}$.

Taking $H_0 \sim$ 500 km s$^{-1}$ Mpc$^{-1}$ and hence our value of $D$ = 5.4x10$^{-28}$ cm$^{-1}$, from the relation $D^2 \sim \kappa\rho$ we obtain instead

$$\rho \sim \frac{D^2}{\kappa} = \frac{D^2 c^2}{8\pi G} = 1.6\text{x}10^{-28}\ \text{g/cm}^3$$

Our estimate of the matter density is two orders of magnitude smaller than Einstein's estimate of 10$^{-26}$ g/cm$^3$ above, but is in good agreement with similar calculations by Eddington (Eddington 1930), Einstein and de Sitter (Einstein and de Sitter 1932) and Einstein some years later (Einstein 1945). Thus it seems that Einstein's estimate of the matter density in the paper under discussion also contains an error.

Corroborating evidence that Einstein's estimates in this paper of the present world radius and matter density of the universe contain an error can be seen in figure 2, a photograph of a blackboard preserved from a lecture given by Einstein at Oxford in April 1931.[17] The equations and calculations of the paper under discussion are shown on the blackboard, while the additional relation $D^2 \sim 10^{-53}$ is also displayed. The latter relation strongly suggests that Einstein erred in his calculation of the quantity *D,* since a Hubble constant of 500 km s$^{-1}$Mpc$^{-1}$ implies a value of $D = 5.4\text{x}10^{-28}$ cm$^{-1}$ or $D^2 = 3\text{x}10^{-55}$ cm$^{-2}$. [18]

---

[16] It is worth noting that our calculation of the quantity $D$ is in good agreement with that of Lemaître (6.8 x 10$^{-28}$ cm$^{-1}$) who used a value of $H_0$ = 625 km s$^{-1}$Mpc$^{-1}$ (Lemaître 1927) and of Einstein in 1945 (4.7 x10$^{-28}$ cm$^{-1}$) when he used a value of $H_0$ = 425 km s$^{-1}$Mpc$^{-1}$ (Einstein 1945).

[17] The occasion was Einstein's second Rhodes lecture at Oxford. A discussion of the lecture can be found in (Eisenberg 2013, p100).

Thus, it seems very likely that Einstein made a numerical error in the calculations of this paper, resulting in an estimate over one hundred times too large for the matter density and twenty times too small for the world radius. That said, the key point is that he tests his model by extracting rough estimates for observable quantities such as the density of matter and the present radius of the universe, so perhaps we should not be too exacting regarding a factor of ten here or there.

## 9. On the age of the universe

It is great interest that Einstein comments on the age problem in this short paper. As noted in section 2, he remarks *"The greatest difficulty of this whole approach is that according to (2 a), the elapsed time since P = 0 comes out at only about $10^{10}$ years".* While the nature of the "difficulty" is not spelt out, the remark is most likely a reference to the fact that the estimated timespan was less than the ages of the stars estimated from astrophysics;[19] other writings of Einstein suggest that he was very aware of this problem.[20]

We note first that it is not obvious how Einstein calculates a timespan of $10^{10}$ years from equation (2a). However, the value coincides exactly with an estimate given by Friedmann in 1922 (Friedmann 1922); as the analysis of Einstein's paper follows Friedmann's work closely, it is very likely that the value is taken from Friedmann. With this in mind, we consider Friedmann's calculation in some detail.

As noted in section 7, Friedmann calculated a timespan for the full evolution of the radius of the universe as:

$$t_\pi = \frac{2}{c}\int_0^{x_0} \sqrt{\frac{x}{A - x + \frac{\lambda}{3c^2}x^3}}\,dx \qquad (6)$$

where A is a constant given by $A = \frac{\kappa M}{6\pi^2}$ and $M$ is the total mass of the universe. This integral has the solution $t_\pi = \frac{\pi A}{c}$ for the case $\lambda = 0$, and taking $M$ as $5\text{x}10^{21}$ solar masses, Friedmann calculated a value of $10^{10}$ years for the timespan of the cycle (Friedmann 1922). However, it has never been clear why Friedmann assumed the value he did for the mass of the universe [21] and the ensuing calculation is in any case problematic. As pointed out by Ari Belenkiy (Belenkiy 2013), an attempt to repeat Friedmann's calculation gives a timespan of

$$t_\pi = \frac{\pi A}{c} = \frac{\kappa M}{6\pi c} = \frac{8GM}{6c^3} = \frac{8(6.67x10^{-11}m^3Kg^{-1}s^{-2})(5x10^{21})(2x10^{30}Kg)}{6(3x10^8ms^{-1})^3} = 3.3x10^{16}s$$

or $1x10^9$ years. This estimate is an order of magnitude smaller than that of Friedmann; thus, Einstein's estimate of the age of the universe appears to repeat a questionable calculation of Friedmann's.[22]

[18] We note that Einstein's value $D^2 \sim 10^{-53}$ cm$^{-2}$ corresponds to a Hubble constant of about 5000 km s$^{-1}$Mpc$^{-1}$.
[19] For example, Condon's estimate of stellar ages of the order of $2\text{x}10^{13}$ years (Condon 1925) was well known
[20] See (Nussbaumer 2014) for example.

Notwithstanding the questionable calculation above, Einstein's attempt to address the problem of a relatively short age for the universe is of great interest. Since many theoreticians addressed the age problem by using the cosmological constant to adjust the timespan of the expansion (Eddington 1930, Lemaître 1934), one is curious to see how Einstein tackles the problem without the term. His solution is that one cannot expect an accurate estimate for the age of the universe given that a fundamental assumption of the theory, that of the homogeneity of matter, is demonstrably invalid. *"One can seek to escape this difficulty by noting that the inhomogeneity of the distribution of stellar material makes our approximate treatment illusory".* Thus Einstein sees the issue as a consequence of the assumptions made in constructing the model; this attitude is quite typical of Einstein's philosophy regarding relativistic cosmology as discussed in the following section.

## 10. On Einstein's philosophy of cosmology

At first sight, Einstein's philosophy in this paper is that of Occam's razor. Confronted with the possibility of an expanding universe, he is willing to remove the cosmological constant from the field equations on the grounds that it is both unsatisfactory (i.e. does not give a stable solution) and apparently redundant. (A year later, he realises that spatial curvature is not a given and removes that also (Einstein and de Sitter 1932)). Einstein's strategy is to construct the simplest possible model of the cosmos, and to test it by extracting estimates of quantities such as the current density of matter and the radius of the universe *"In any case, this theory is simple enough that it can be easily compared with the astronomical facts"* (see section 2). Thus, Einstein does not discuss the details of the matter-energy tensor in this paper, nor does he ask how galaxies might form in an expanding universe, a major question in dynamic models.

---

[21] See for example (Tropp, Frenkel and Chernin 1993) p 159-161.
[22] Indeed, it can be shown that a universe of closed geometry and $\lambda = 0$ must have an age less than $2/(3H_0)$, i.e., $10^9$ billion years for a Hubble constant of 500 km $s^{-1}Mpc^{-1}$.

There is much to be admired in this minimalist, pragmatic approach. It should not be forgotten that it was by no means yet accepted that Hubble's observations truly constituted evidence for an expanding universe. Thus Einstein's approach is one of exploration, rather than an abrupt shift to a new paradigm. It is of great interest that the paper under discussion contains a significant number of caveats. For example, we saw in section 2 that the statement *"Now that it has become clear from Hubbel's* [sic] *results"* appears with the important qualifier *"(at least if their systematic redshifts are to be interpreted as Doppler effects)"*. Likewise, Einstein's conclusions at the end of the paper are mitigated by the statement *"It further shows how careful one must be with large extrapolations over time in astronomy"* (see Appendix). Thus, the article offers a revealing glimpse of the process of change from one paradigm to another, and this glimpse does not suggest an abrupt transition to a new paradigm that becomes incommensurate with its predecessor, as proposed by Thomas Kuhn.[23] Instead, Einstein is exploring how well a cosmic model without a cosmological constant matches certain tentative new observations, a much more nuanced approach. It is worth emphasizing once again that the starting point for the model of this paper is the instability of the static solution, not Hubble's observations.

It is intriguing that the phenomenological approach of this paper is more typical of the young Einstein than of the rather mathematical formalism of his other works around this time (Einstein and Mayer 1931a, 1931b). We suggest that this reflects the topic Einstein is addressing. Although it is not yet clear that Hubble's observations truly constitute evidence for an expanding universe, there is a real prospect that the data represent an astonishing new phenomenon; thus Einstein's first paper on the subject is very typical of his pragmatic approach to emerging phenomena (Einstein 1905a, b).

On the other hand, it must be admitted that there is some evidence of haste in Einstein's paper. We have already noted in section 2 the lack of references to key papers in relativistic cosmology, both before and after Hubble's graph of 1929. In particular, it seems a pity that there is no recognition of Lemaître's early proposal of a relativistic universe of expanding radius as an explanation for preliminary measurements of the redshifts and distances of the spiral nebulae (Lemaître, 1927),[24] a proposal Einstein was unquestionably aware of.[25] Einstein's attitude towards observational cosmology seems equally casual; given the central importance of Hubble's observations to his model, it is strange that there is no specific citation of the seminal 1929 article (Hubble 1929). More importantly, Einstein seems unaware that Hubble's 1929 graph relied heavily on the pioneering redshift observations of the astronomer Vesto Slipher;[26] indeed, Einstein's exclusive reference to Hubble (here and elsewhere) may have contributed to the eclipsing of Slipher's contribution to the discovery of the expanding universe (O'Raifeartaigh 2013, Nussbaumer 2013, Way 2013).

[23] The concept of incommensurability refers to Kuhn's belief that a new scientific paradigm cannot be meaningfully compared with previous models, because the underlying assumptions of the worldviews are different (Kuhn, 1962).

Taking together the lack of references to previous work, the close tracking of Friedmann's analysis and the anomalies in the calculations of the density, radius and timespan of the universe, Einstein's paper seems something of a quick fix. One is not surprised to learn that it took only four days to write, as noted in our introduction. Thus, Einstein seems something of an impatient cosmologist, rather than a scientist attempting to show that his greatest theory may be compatible with some astonishing new astronomical observations. This impression is strengthened by the fact that he did not follow the paper with a comprehensive overview, but with an even shorter article one year later (Einstein and de Sitter, 1932). After this point, Einstein occupied himself very little with cosmology, [27] despite the exciting work of Lemaître in the 1930s and Gamow in the 1940s, both of whom he knew personally. This indifference seems extraordinary given that the tentative observations of a cosmic expansion represented an astounding phenomenon that could be readily understood in relativistic terms.

Several possible reasons for Einstein's ensuing lack of interest in cosmology suggest themselves. The first is that one must recall once more that it was not yet established beyond doubt that Hubble's observations truly constituted evidence of an expanding universe. As well as the caveats of the paper under discussion, a certain circumspection can be discerned in Einstein's other writings in 1931: *"New observations by Hubble and Humason concerning the redshift of light in distant nebulae make the presumptions near that the general structure of the universe is not static*" (AP 1931a) and: *"The red shift is still a mystery…and might be interpreted as the light quanta getting redder by losing energy as they went long distances"* (AP 1931b).

[24] Some recent reviews of Lemaître's contribution can be found in (Holder and Mitton 2013).

[25] Lemaître himself had apprised Einstein of the result in 1927; Einstein found Lemaître's expanding universe 'abominable' while Lemaître felt that Einstein was not aware of emerging astronomical results (Lemaître 1958).

[26] Most of the redshifts in Hubble's graph of 1929 are from Slipher (O'Raifeartaigh 2013).

[27] An exception is the essay 'On the Cosmological Problem', included as an appendix to the book *'The Meaning of Relativity'* from 1945 onwards (Einstein 1945).

A more profound reason may be that Einstein simply felt more comfortable with the notion of a static, bounded universe from a philosophical point of view. That this was his preferred option is clear from his negative reaction to the dynamic models of Friedmann and Lemaître when they were first mooted.[28] Is it possible that, faced with the possibility of a dynamic universe, Einstein simply lost confidence in his own intuition? Another question was whether one could have confidence in relativity at such enormous scales, or indeed whether one could truly construct realistic models of the cosmos from mathematics.[29]

With the emergence of Hubble's results Einstein may be intrigued, but he immediately notes that if the phenomenon truly represents an expansion of space, it raises difficult questions concerning both the age and the origin of the universe. It is interesting that Einstein finds room to comment on both the issue of the singularity and the age problem in this short paper, while Friedmann does not consider the problem of the singularity in his more detailed analysis, nor does he attempt compare his age estimate with stellar ages (Friedmann 1922, 1924).

Einstein's remark in this paper on the issue of the singularity may be an important clue in discerning his general attitude towards relativistic cosmology. As pointed out in section 6 above, he clearly viewed the issue as a shortcoming of theory. This view is consistent with Einstein's view of singularities in general relativity throughout the 1920s as a limitation of theory that might one day be overcome in a more general framework (Earman and Eisenstaedt 1999). Einstein's view is best summed up in his 1945 essay: "*The present relativistic theory of gravitation is based on a separation of the concepts of 'gravitational field' and of 'matter'. It may be plausible that the theory is for this reason inadequate for very high density of matter. It may well be the case that for a unified theory there would arise no singularity*" (Einstein 1945, p 124). Thus, we suggest that Einstein's view of general relativity as incomplete, and his intense focus on the search for a more complete unified field theory,[30] may have blinded him somewhat to the foundational role relativity had to play in the new cosmology.

**11. Conclusions**

Einstein's paper "*On the cosmological problem of the general theory of relativity*" (Einstein 1931) is not widely known but is of historical interest because it offers many insights into Einstein's thinking on cosmology in the light of the first tentative evidence for an expanding universe, and because it contains a model of the cosmos distinct from his static model of 1917 or the Einstein-de Sitter model of 1932.

[28] Einstein's reaction to each is described in the essay by Harry Nussbaumer in this volume.
[29] This point is considered in (Eisenstaedt 1989).
[30] See (Pais, chapter 17).

Einstein's principle aim in the paper is to show that relativity can account for a universe of expanding radius if the cosmological constant is removed from the field equations. Following Friedmann's analysis of 1922, he constructs a model universe of time-varying radius, finite duration and positive curvature in which the universe first expands and then contracts. However, the model is not defined at the endpoints and we suggest that it is not accurate to describe the model as periodic or cyclic.

Einstein's approach to cosmology can be understood in terms of the philosophy of Occam's razor. His strategy is to investigate whether the simplest relativistic model can account for the facts. To validate his model, Einstein extracts estimates for the density of matter, the radius of the universe and the timespan of the expansion and compares them with observations. However, these calculations contain some anomalies.

The cosmic model of this paper was soon superseded by a revised model in which the assumption of spatial curvature was removed (Einstein and deSitter 1932). Many years later, it seems the assumption of a vanishing cosmological constant may also have been unjustified.[31] However, the 1931 paper casts some light on Einstein's view of emerging cosmological problems such as the singularity and the timespan of the expansion. We suggest that he sees such problems as limitations of the general theory of relativity, a view that may explain his rather hasty approach in the paper and his apparent lack of interest in relativistic cosmology in the years to come.

## Acknowledgements

The authors would like to thank the Albert Einstein Archives of the Hebrew University of Jerusalem for permission to publish our translation of Einstein's article (see Appendix). We would also like to thank the Dublin Institute of Advanced Studies for access to the Collected Papers of Albert Einstein (Princeton University Press), the Oxford Museum for the History of Science for the use of a photograph of Einstein's blackboard and Professor Andrew Liddle of Edinburgh University for helpful suggestions.

[31] See (Solà 2013) for an overview of dark energy and the cosmological constant.

## References

Associated Press Report.1931a. Prof. Einstein begins his work at Mt. Wilson. *NYT*, Jan 3, p1

Associated Press Report. 1931b. Redshift of nebulae a puzzle, says Einstein. *NYT*, Feb 12, p2

Bartusiak, M. 2009. *The Day We Found the Universe* .Vintage Books, New York

Belenkiy, A. 2012. *Physics Today* **65** (10): 38-43

Belenkiy, A. 2013. The waters I am entering no one yet has crossed: Alexander Friedmann and the origins of modern cosmology. *Origins of the Expanding Universe: 1912-1932*, edited by M. Way and D. Hunter, Astron. Soc. Pac. Conference Series **471:** 71-96

Condon, E. (1925). The ages of the stars. *Proc. Nat. Acad. Sci.* **11**: 125-130

de Sitter, W. 1917. On Einstein's theory of gravitation and its astronomical consequences. *Month. Not. Roy. Astron. Soc*.**78:**3-28

de Sitter, W. 1930a. On the magnitudes, diameters and distances of the extragalactic nebulae , and their apparent radial velocities. *Bull. Astron. Inst. Neth.* **185**: 157-171

de Sitter, W. 1930b. The expanding universe. Discussion of Lemaitre's solution of the equations of the inertial field. *Bull. Astron. Inst. Neth***. 193**: 211-218

Vibert Douglas, A. 1956. *The Life of Arthur Stanley Eddington*. Nelson and Sons Ltd.

Eddington, A.S.1930.On the instability of Einstein's spherical world. *Month. Not. of the Roy. Astron. Soc.* **90**, 668-678

Eddington A.S.1931. The recession of the extra-galactic nebulae. *Proc. Roy. Soc.* A **133**, 3-10

Einstein, A. 1905a. Zur Elektrodynamik bewegter Körper. *Annal. Physik* **17**: 891-921 reprinted and translated in *The Principle of Relativity* (Dover, 1952) p 35-65

Einstein, A. 1905b. Über einen die Erzeugung und Verwandlung des Lichtes betreffenden heuristischen Gesichtspunkt *Annal. Physik* **17**: 132-148

Einstein, A. 1917. Kosmologischen Betrachtungen zur allgemeinen Relativitätstheorie *Sitzungsb. König. Preuss. Akad*. 142-152, reprinted and translated in *The Principle of Relativity* (Dover, 1952) p 175-188

Einstein, A. 1923. Letter to H. Weyl, May 23

Einstein, A. 1931. Zum kosmologischen Problem der allgemeinen Relativitätstheorie *Sitzungsb.König. Preuss. Akad*. 235-237

Einstein, A. 1934. Review of *Relativity, Thermodynamics and Cosmology* (Oxford University Press) by Richard Tolman *Science* **80**: 358

Einstein, A. 1945. On the cosmological problem. Appendix to *The Meaning of Relativity*. Princeton University Press, Princeton, New Jersey, 2nd ed. 109-132. Also available in later editions

Einstein, A. and W. Mayer. 1931a. Systematische Untersuchung über kompatible Feldgleichungen welche in einem Riemannschen Raume mit Fern-Parallelismus gesetzt werden können *Sitzungsb. König. Preuss. Akad.* 257–265

Einstein, A. and W. Mayer. 1931b. Einheitliche Feldtheorie von Gravitation und Elektrizität . *Sitzungsb. König. Preuss. Akad.* 541–557

Einstein, A. and W. de Sitter, W. 1932. On the relation between the expansion and the mean density of the universe *Proc. Nat. Acad. Sci.* **18**: 213-214

Earman, J. and J. Eisenstaedt. 1999. Einstein and Singularities *Stud. Hist.Phil.Mod. Phys.***30(2)**:185-235

Eisenstaedt, J. 1989. 'The low water-mark of general relativity 1925-1950' in *Einstein and the History of General Relativity, Einstein Studies vol 1,* edited by J. Stachel and D. Howard. Birkhauser, Boston, Vol. 1, p277-292

Eisenstaedt, J .2006. *The Curious History of Relativity*: *How Einstein's Theory was Lost and Found Again.* Princeton University Press

Eisinger, J .2011. *Einstein on the Road.* Prometheus Books, New York

Friedman, A. 1922. Über die Krümmung des Raumes. *Zeit. Physik.* **10** : 377-386

Friedman, A. 1924. Über die Möglichkeit einer Welt mit constanter negative Krümmung des Raumes. *Zeit.Physik.* **21**: 326-332

Gamow, George. 1956. The Evolutionary Universe. *Sci. Amer.* **195** (3):136-156

Holder, R.D. and S. Mitton. 2013. *Georges Lemaître: Life, Science and Legacy*. Series: Astrophysics and Space Science Library, **395:**XII Springer Verlag

Hubble, E. 1929. A relation between distance and radial velocity among extra-galactic nebulae. *Proc. Nat. Acad. Sci.* **15**: 168-173

Kuhn, T. (1962) *The Structure of Scientific Revolutions*. Cambridge University Press, Cambridge

Lemaître, G. 1927. Un univers homogène de masse constant et de rayon croissant, rendant compte de la vitesse radiale des nébeleuses extra-galactiques. *Annal. Soc. Scien. Brux.* Série A. **47:** 49-59

Lemaître, G. 1934. Evolution of the expanding universe. *Proc. Nat. Acad. of Sci.* **20:** 12-17

Lemaître, G. 1958. Rencontres avec Einstein. *Rev. Quest. Scien.* **129:** 129-132

Livio, M. 2013. *Brilliant Blunders: From Darwin to Einstein - Colossal Mistakes by Great Scientists That Changed Our Understanding of Life and the Universe.* Simon and Schuster.

North, J.D. 1965. *The Measure of the Universe: A History of Modern Cosmology* . Oxford University Press

Nussbaumer, H. and L. Bieri. 2009. *Discovering the Expanding Universe*, Cambridge University Press, Cambridge

Nussbaumer, H. 2013. Slipher's redshifts as support for de Sitter's model and the discovery of the dynamic universe. *Origins of the Expanding Universe: 1912-1932*, edited by M. Way and D. Hunter, Astron. Soc. Pac. Conference Series **471:** 25-38

Nussbaumer, H. 2014. Einstein's conversion from his static to an expanding universe. *Eur. Phys. J. (H),* this issue.

O'Raifeartaigh, C. 2013. The contribution of V.M. Slipher to the discovery of the expanding universe. *Origins of the Expanding Universe: 1912-1932*, edited by M. Way and D. Hunter, Astron. Soc. Pac. Conference Series **471:** 49-62

Pais, A. 1982. *Subtle is the Lord: The Science and the Life of Albert Einstein*, Oxford University Press, Oxford

Peacock, J. A. 2013. Slipher, galaxies and cosmological velocity fields. *Origins of the Expanding Universe: 1912-1932*, edited by M. Way and D. Hunter, Astron. Soc. Pac. Conference Series **471:** 3-25

Peebles, P.J. and Ratra, Bharat. 2003. The cosmological constant and dark energy. *Rev. Mod.Phys* **75(2):** 559-606

Rindler, W. 1969. *Essential Relativity: Special, General and Cosmological.* Van Nostrand Reinhold Company New York. Reprinted 1977 Springer-Verlag New York 2nd ed. Reprinted 2006 Oxford University Press.

Robertson, H. 1929. On the foundations of relativistic cosmology. *Proc. Nat. Acad. of Sci.* **15:** 822-829

Solà, J. 2013. Cosmological constant and vacuum energy: old and new ideas. *J. Phys. Conference Series* **453:** (1) 012015

Steinhardt, P.J. and N. Turok 2007. *Endless Universe: Beyond the Big Bang*. Weidenfeld and Nicolson, London

Tolman, R.C. 1929. On the possible line elements for the universe. *Proc. Nat. Acad. of Sci*. **15:** 297-304

Tolman, R.C. 1930. More complete discussion of the time-dependence of the non-static line element for the universe. *Proc. Nat. Acad. Sci.* **16:** 409-420

Tolman, R.C. 1931. On the theoretical requirements for a periodic behaviour of the universe. *Phys. Rev.* **38:** 1758-1771

Tolman, R.C. 1932. On the behaviour of non-static models of the universe when the cosmological term is omitted. *Phys. Rev.* **39:** 835-843

Tropp, E.A., Frenkel V.Y. and Chernin A.D. 1993. *Alexander A. Friedmann: The Man Who Made The Universe Expand* . Cambridge University Press, Cambridge

Way, M. 2013. Dismantling Hubble's Legacy? *Origins of the Expanding Universe: 1912-1932*, edited by M. Way and D. Hunter, Astron. Soc. Pac. Conference Series **471:** 97-132

Weyl, H. 1923. Zur allgemeinen Relativitätstheorie. *Physik. Zeitschrift*. **24:** 230-232

## Figure 1

Einstein's diagram of the evolution of the world radius for the case of positive spatial curvature and λ = 0 (Einstein 1945). Reproduced by permission of Princeton University Press, Princeton, New Jersey.

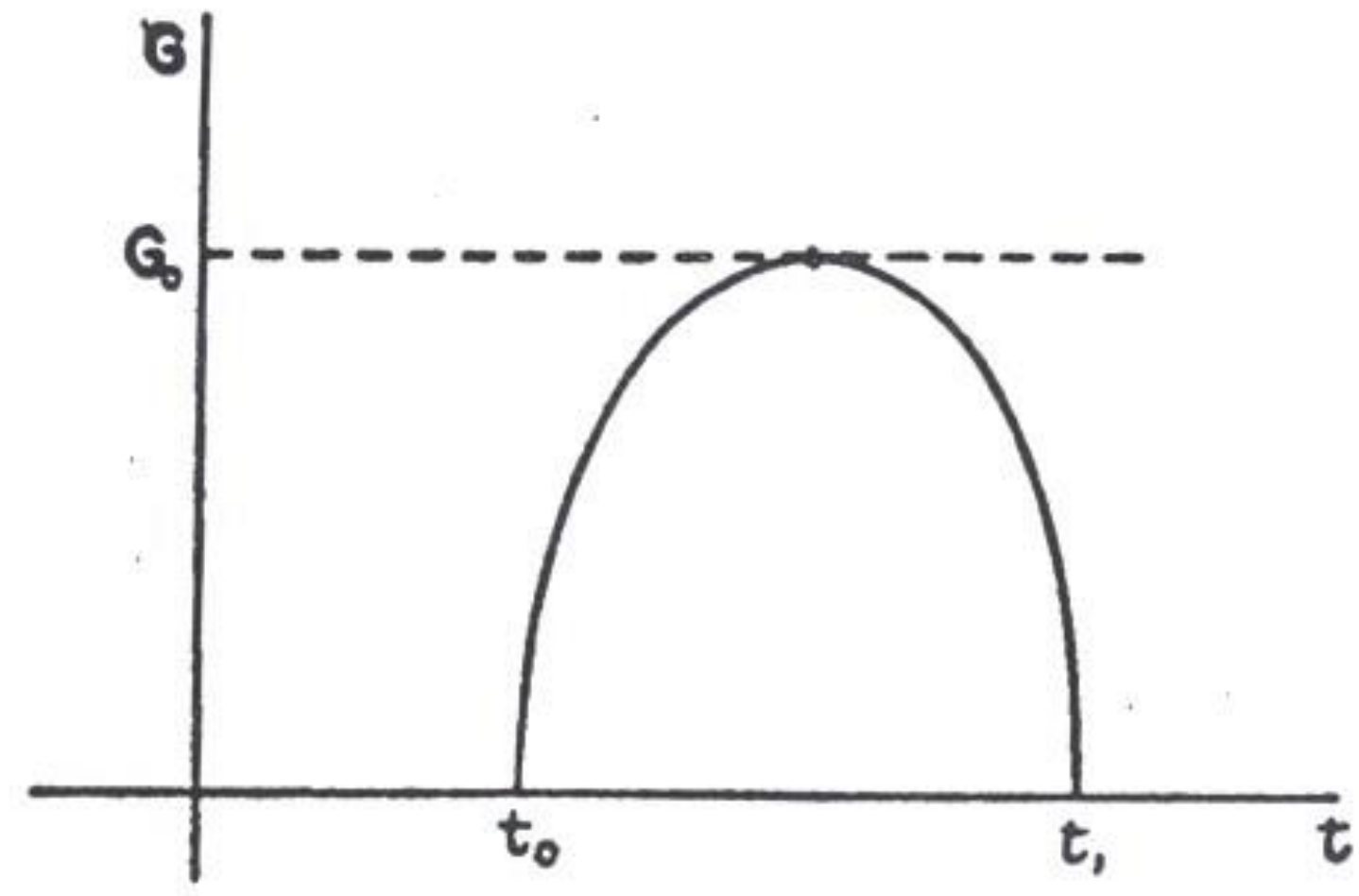

## Figure 2

An image of the blackboard used in Einstein's 2nd Rhodes lecture at Oxford in April 1931 (reproduced by permission of the Museum of the History of Science, University of Oxford). The equations shown are taken directly from the paper under consideration: the extra equation $D^2 \sim 10^{-53}$ suggests that Einstein made a numerical error in his use of the Hubble constant to calculate estimates for the mass density and radius of the universe.

$$D = \frac{1}{c}\frac{1}{l}\frac{dl}{dt} = \frac{1}{c}\frac{1}{P}\frac{dP}{dt}$$

$$D^2 = \frac{1}{P^2}\frac{P_0 - P}{P} \sim \frac{1}{P^2} \qquad (1a)$$

$$D^2 = \frac{\kappa\rho}{3}\frac{P_0 - P}{P_0} \sim \kappa\rho \qquad (2a)$$

$$D^2 \sim 10^{-53}$$

$$\rho \sim 10^{-26}$$

$$P \sim 10^8 \text{ L.J.}$$

$$t \sim 10^{10}\,(10^{11})\text{ J}$$

**Appendix**

## ON THE COSMOLOGICAL PROBLEM OF THE GENERAL THEORY OF RELATIVITY*

by

A. EINSTEIN**

The cosmological problem is understood to concern the question of the nature of space and the manner of the distribution of matter on a large scale, where the material of the stars and stellar systems is assumed for simplicity to be replaced by a continuous distribution of matter. Since I began to address this problem, shortly after advancing the general theory of relativity, not only have numerous theoretical articles on the subject appeared, but facts have come to light from Hubbel's investigations of the Doppler effect and the distribution of extra-galactic nebulae that indicate new directions for the theory.

In my original investigation, I proceeded from the following assumptions:

1. All locations in the universe are equivalent; in particular the locally averaged density of stellar matter should therefore be the same everywhere.
2. Spatial structure and density should be constant over time.

At that time, I showed that both assumptions can be accounted for with a non-zero mean density $\rho$, if the so-called cosmological term is introduced into the field equations of the general theory of relativity such that these read:

$$\left(R_{ik} - \frac{1}{2} g_{ik} R\right) + \lambda g_{ik} = -\kappa T_{ik} \quad \ldots.. \quad (1)$$

These equations can be satisfied by a spatially spherical static world of radius $P = \sqrt{\frac{2}{\kappa\rho}}$ where $\rho$ is the (pressure-free) mean density of matter.

Now that it has become clear from Hubbel's results that the extra-galactic nebulae are uniformly distributed throughout space and are in dilatory motion (at least if their systematic redshifts are to be interpreted as Doppler effects), assumption (2) concerning the static nature of space has no longer any justification and the question arises as to whether the general theory of relativity can account for these findings.

*Translated by C. O'Raifeartaigh and B.McCann. The original article appeared as *Zum kosmologischen Problem der allgemeinen Relativitätstheorie von A. Einstein* in Sonderasugabe aus den Sitzungsb.König. Preuss. Akad. Wissen., Phys.-Math Klasse. 1931. XII

**Deceased

Several investigators have attempted to account for the new facts by means of a spherical space, whose radius $P$ is variable over time. The first to try this approach, uninfluenced by observations, was A. Friedman,[1] on whose calculations I base the following remarks. One proceeds accordingly from a line element of the form

$$ds^2 = -P^2(dx_1^2 + sin^2 x_1\, dx_2^2 + sin^2 x_1 sin^2 x_2 dx_3^2) + c^2 dx_4^2 \;.... \qquad (2)$$

where $P$ is understood to be a function of the real-valued time variable $x_4$ alone. For the determination of $P$ and the relation of this quantity to the (variable) density $\rho$ he derives from (1) the two differential equations

$$\frac{P'^2}{P^2} + \frac{2P''}{P} + \frac{c^2}{P^2} - \lambda = 0 \;.... \qquad (2)$$

$$\frac{3P'^2}{P^2} + \frac{3c^2}{P^2} - \lambda = \kappa c^2 \rho \;.... \qquad (3)$$

From these equations, one obtains my previous solution if it is assumed that $P$ is constant over time. However, it can also be shown with the help of these equations that this solution is not stable, i.e., a solution that deviates only slightly from the static solution at a given point in time will differ ever more from it with the passage of time. On these grounds alone, I am no longer inclined to ascribe a physical meaning to my former solution, quite apart from Hubbel's observations.

Under these circumstances, one must ask whether one can account for the facts without the introduction of the λ-term, which is in any case theoretically unsatisfactory. We consider here to what extent this is the case, neglecting, like Friedman, the effects of radiation. As Friedman has shown, it follows from (2) by integration that (for λ = 0)

$$(\frac{dP}{dt})^2 = c^2 \frac{P_0 - P}{P} \;, ....... \qquad (2a)$$

where $P_0$ denotes a constant of integration that sets an upper bound for the world radius that cannot be exceeded with the passage of time. At this point a change of sign for $\frac{dP}{dt}$ must occur [2]. From (3), it follows that $\rho$ (for λ = 0) will in any case be positive, as it must be.

It follows from Hubbel's results that for the present, it is to be assumed that $\frac{dP}{dt} > 0$, and that the Doppler effect divided by the distance is a quantity that is independent of distance, which here can be expressed with sufficient accuracy by the quantity $D = \frac{1}{P}\frac{dP}{dt}.\frac{1}{c}$. Instead of equation (2a) one can write

1 Zeitschr. f. Physik. 10. p 377. I922.

2 According to (2a) P cannot grow beyond *Po* and according to (2),*P* cannot remain at the value of $P_0$.

$$D^2 = \frac{1}{P^2}\frac{P_0 - P}{P} \dots \qquad (2b)$$

and instead of (3)

$$D^2 = \frac{1}{3}\kappa\rho\frac{P_0 - P}{P_0} \dots . \qquad (3a)$$

The process described by (2a) is as follows. For small $P$ (our idealization fails for the strict limiting case $P = 0$), $P$ grows very rapidly. Thereafter the rate of change $\frac{dP}{dt}$ decreases ever more with increasing $P$, and disappears when the limit $P = P_0$ is attained, whereupon the whole process is gone through in the opposite sense (i.e. with ever rapidly decreasing $P$).

If we want to compare our formulas with the facts, we must assume that we are somewhere in the phase of increasing $P$. For a rough orientation, it is probably reasonable to assume that $P - P_0$ is of the same order of magnitude as $P_0$, so that we obtain solely from the order of magnitude the equation

$$D^2 \sim \kappa\rho$$

which gives an order of magnitude of $10^{-26}$ for the density $\rho$, which seems to be somewhat in agreement with the estimates of the astronomers. Similarly, from the order of magnitude, the present world radius is determined in accordance with equation (2b) by

$$P \sim \frac{1}{D}$$

which however gives a value of only about $10^8$ light-years.

However, the greatest difficulty with the whole approach, as is well-known, is that according to (2 a), the elapsed time since $P = 0$ comes out at only about $10^{10}$ years. One can seek to escape this difficulty by noting that the inhomogeneity of the distribution of stellar material makes our approximate treatment illusory. Moreover, it should be noted that it is unlikely that any theory that interprets Hubbel's tremendous shifts of spectral lines as Doppler effects will easily avoid this difficulty.

In any case, this theory is simple enough that it can be easily compared with the astronomical facts. It further shows how careful one must be with large extrapolations over time in astronomy. Most importantly, it seems that the general theory of relativity can account for Hubbel's new facts more naturally (namely, without the λ-term), than it can the postulate of the quasi-static nature of space, which has now been rendered a remote possibility empirically.

## Typographical notes from the translators

(i) We have preserved the same page layout as the original article in terms of paragraph structure, number of pages and numbering for equations.

(ii) In the original article, equation (1) on the first page appears as

$$\left(R_{i\kappa} - \frac{1}{2} g_{i\kappa} R\right) + \lambda g_{i\kappa} = -k T_{i\kappa}$$

As the tensor indices in this equation were usually written in lower case letters and the Einstein constant as $\kappa$ (Einstein, 1917, Friedmann 1922, Einstein 1945), we have rewritten equation (1) slightly as

$$\left(R_{ik} - \frac{1}{2} g_{ik} R\right) + \lambda g_{ik} = -\kappa T_{ik}$$

(iii) In the original article, the relation immediately after equation (1) appears as $P = \sqrt{\frac{2}{\chi\rho}}$. The symbol $\chi$ is not defined and also appears in equations (3) and (3a). As it is clear from the context that this symbol represents the Einstein constant $\kappa$ (Einstein 1917, Friedmann 1922, Einstein 1945), we have replaced the symbol $\chi$ with $\kappa$ each time it appears in the article.

(iv) Two distinct equations are labelled as equation (2) on the second page of the article. When Einstein refers to equation (2) in the text, it is clearly the differential equation he means.

(v) The name Hubble is misspelt as Hubbel each time it occurs in the article. We have retained this misspelling for authenticity and it may be of some significance as discussed in the notes above.

(vi) Units of measurement are not given for the estimate $\rho = 10^{-26}$ on the third page of the article. In all other known instances, Einstein used units of g/cm$^3$ for this quantity in accordance with standard usage at the time (Einstein 1932, 1945).

(vii) The symbol $\sim$ in the last two equations of the paper appears in the published paper as $\infty$. From the context, the symbol is clearly intended to mean 'approximately equal to'. Following Einstein's lead in figure 2, we have rewritten this symbol as $\sim$ .